\documentclass[superscriptaddress, reprint, amsmath, amssymb, aps, pra, floatfix]{revtex4-2}
\usepackage{graphicx} 

\usepackage{xcolor} 
\usepackage{hyperref} 
\usepackage{orcidlink} 
\hypersetup{
	colorlinks=true,
	linkcolor=blue,
	filecolor=magenta,
	urlcolor=magenta,
}

\definecolor{dgreen}{rgb}{0.0,0.5,0.0}
\definecolor{pink}{rgb}{1,0,0.9}
\newcommand{\fixme}[1]{{\color{pink}#1}}
\usepackage[normalem]{ulem}

\begin{document}
\title{Magnetic phases in the $J_{1}$-$J_{2}$ antiferromagnetic XY model\\ on the honeycomb lattice }

\author{I.V. Lukin\orcidlink{0000-0002-8133-2829}}
\email{illya.lukin11@gmail.com}
\affiliation{Akhiezer Institute for Theoretical Physics, NSC KIPT, Akademichna 1, 61108 Kharkiv, Ukraine}
\affiliation{Haiqu, Inc., 95 Third Street, San Francisco, CA 94103, USA}

\author{M.O. Luhanko\orcidlink{0000-0002-6478-0419}}
\affiliation{Haiqu, Inc., 95 Third Street, San Francisco, CA 94103, USA}
\affiliation{Education and Research Institute ``School of Physics and Technology'', Karazin Kharkiv National University, Svobody Square 4, 61022 Kharkiv, Ukraine}

\author{Yu.V. Slyusarenko\orcidlink{0000-0001-5298-0731}}
\affiliation{Akhiezer Institute for Theoretical Physics, NSC KIPT, Akademichna 1, 61108 Kharkiv, Ukraine}
\affiliation{Education and Research Institute ``School of Physics and Technology'', Karazin Kharkiv National University, Svobody Square 4, 61022 Kharkiv, Ukraine}

\author{A.G. Sotnikov\orcidlink{0000-0002-3632-4790}}
\email{a\_sotnikov@kipt.kharkov.ua}
\affiliation{Akhiezer Institute for Theoretical Physics, NSC KIPT, Akademichna 1, 61108 Kharkiv, Ukraine}
\affiliation{Education and Research Institute ``School of Physics and Technology'', Karazin Kharkiv National University, Svobody Square 4, 61022 Kharkiv, Ukraine}

\begin{abstract}
    We study ground-state properties and phase diagram of the $J_{1}$-$J_{2}$ antiferromagnetic XY model on the honeycomb lattice by means of the developed corner transfer matrix renormalization group algorithm with the two-site unit cell and the infinite spiral projected entangled pair states ansatz. 
    We identify the main phases: N\'{e}el, Ising, collinear, and incommensurate spiral phases, as well as the transitions between them, as functions of the ratio $J_{2}/J_{1}$.
    In the regime of competing types of ordering, we show that the energies of the dimerized states are systematically higher than the energies in the collinear phase. This collinear phase transforms to the incommensurate spiral phase through the second-order phase transition upon a further increase of $J_2/J_1$.
\end{abstract}

\date{\today}

\maketitle

\section{Introduction}

Frustrated spin systems are known to host unusual correlated phases as quantum spin liquids, valence bond solids, spiral magnetic orders and exotic commensurate magnetic orders~\cite{lacroix2011introduction}. The spin $S=1/2$ XY-model with nearest-neighbor (NN) and next-nearest-neighbor (NNN) interactions is a spectacular example of a frustrated system, with all of the above-mentioned exotic phases being strong contenders to appear in its phase diagram. In particular, in different studies it was predicted to host Bose metal phase~\cite{varney2011kaleidoscope}, exotic magnetic order with $Z$-directed magnetization \cite{zhu2013unexpected}, spiral phase~\cite{DiCiolo2014} and valence bond solid phase~\cite{zhu2013unexpected, zhu2014quantum, bishop2014frustrated}. Methods such as exact diagonalization~\cite{varney2011kaleidoscope}, density-matrix renormalization group (DMRG) \cite{zhu2013unexpected, zhu2014quantum, huang2021quantum}, series expansion~\cite{Oitmaa2014}, coupled cluster method~\cite{bishop2014frustrated} and variational Monte Carlo \cite{carrasquilla2013nature, DiCiolo2014} were used to predict the ground states of the model at different couplings. Still, its phase diagram is not yet completely settled, with the new findings on the subject continuing to appear~\cite{satoori2023quantum, chernyshev2025demystifying}. In this study, we add projected entangled pair states (PEPS) to this list of methods. 

Projected entangled pair states \cite{Verstraete2004, Verstraete2008, Orus2014, Cirac2021} are an example of tensor network states \cite{Orus2019, Banuls2023} and represent a direct two-dimensional generalization of matrix product states~\cite{Schollwock2011}, which are the basis of the celebrated DMRG algorithm \cite{White1993}. A successful application of PEPS is enabled by the area law principle \cite{Eisert2010}, which bounds the subregion entanglement by the subregion boundary area. The PEPS states represent precisely the states obeying this bound. The recently-proposed spiral PEPS ansatz~\cite{hasik2024incommensurate} generalizes PEPS states to represent spiral magnetic orders; it was also recently employed in Ref.~\cite{Schmoll2024}. 

The central object of this study is the infinite lattice spiral PEPS state (iPEPS) with the two-site quasi-periodic unit cell (periodic up to spiral rotation). To optimize these states, we employ the variational (gradient-based) approach, as was originally proposed in Refs.~\cite{Corboz2016, Vanderstraeten2016}. We obtain gradients for optimization with automatic differentiation \cite{Liao2019, Naumann2024}, while we compute all the observables, including the energy (necessary in optimization) with the version of the corner transfer matrix renormalization group (CTMRG) \cite{Nishino1996, Nishino1997, Orus2009} on the honeycomb lattice \cite{Lukin2023, Nyckees2023, Gendiar2012}, which is specifically developed in this paper. 
Note that CTMRG algorithms on the honeycomb and triangular lattices were employed previously in a similar context of variational PEPS optimizations~\cite{Naumann2025, Yang2026, nyckees2026tensor, ghosh2025simplex, ghosh2025symmetry, zhang2025topological}. Using this approach, we optimize PEPS states for different regimes in the model parameter space and construct the phase diagram.

\section{Model}

\begin{figure}
    \includegraphics[width=\linewidth]{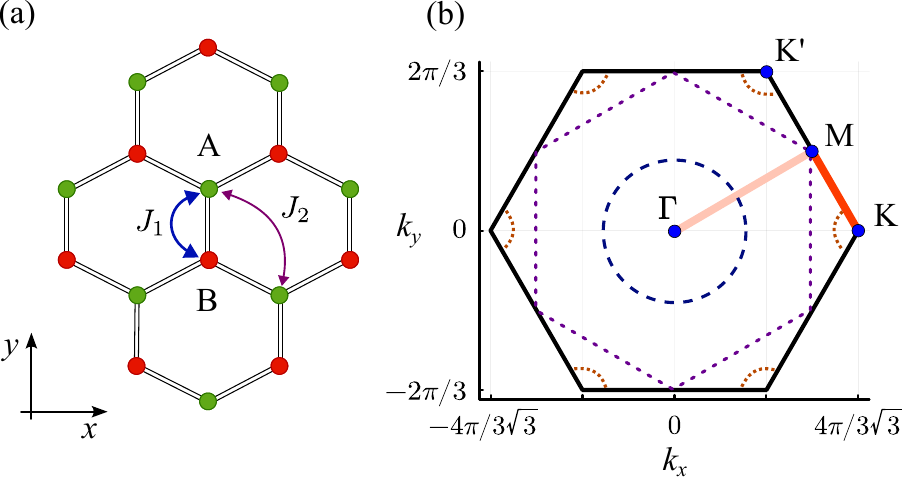}
    \caption{(a) Honeycomb lattice with the two-site unit cell in the real-space representation with magnetic couplings $J_1$ and $J_2$. (b) Brillouin zone with high-symmetry points, some manifolds for classically degenerate spiral states (dotted lines), and highlighted lines $\mathbf{\Gamma}$-$\mathbf{M}$ and $\mathbf{M}$-$\mathbf{K}$, where one can expect stability of the spiral states according to the $1/S$ expansion~\cite{DiCiolo2014}.}
    \label{fig:1}
\end{figure}

We focus on the spin-1/2 antiferromagnetic XY model with the nearest-neighbor and next-nearest neighbor interactions on the honeycomb lattice. The model Hamiltonian is defined as follows:
\begin{equation}\label{eq:XY_model}
    H = J_{1} \sum_{\langle ij \rangle} \left(S^{x}_{i} S^{x}_{j} + S^{y}_{i} S^{y}_{j}\right) 
    + J_{2} \sum_{\langle \langle ij \rangle \rangle} \left(S^{x}_{i} S^{x}_{j} + S^{y}_{i} S^{y}_{j}\right), 
\end{equation}
where $\langle ij\rangle$ and $\langle \langle ij \rangle \rangle$ are the NN and NNN pairs of spins on the honeycomb lattice, respectively, as illustrated in Fig.~\fixme{\ref{fig:1}}, while $S^{x}_{i}$ and $S^{y}_{i}$ are the conventional spin-$1/2$ operators on the sites of the lattice. We take both couplings $J_{1}$ and $J_{2}$ as antiferromagnetic (positive) to frustrate spin orientations on the lattice. Note that the model can be equivalently considered as a system of hard-core bosons with NN and NNN hopping processes. 

At the classical level (in the limit of spin $S\to\infty$), the phase diagram of the model is known in good detail~\cite{DiCiolo2014}. For $J_{2}/J_{1} < 1/6$, the model is in the antiferromagnetic N\'{e}el phase with spins oriented exclusively in the $XY$ plane. At larger values of $J_{2}/J_{1}$, the model has a degenerate manifold of classical ground states formed by spiral magnetic orders with different spiral wave vectors~$\mathbf{q}$. For $J_{2}/J_{1} < 1/2$, these degenerate spiral states have wave vectors near the $\Gamma$ point in the Brillouin zone (near the N\'{e}el order), as shown in Fig.~\ref{fig:1}. 
At $J_{2}/J_{1}=1/2$, the spiral states may have a wave vector $\mathbf{q}=\mathbf{M}$ in the Brillouin zone, corresponding to the collinear magnetic order. At even higher values of $J_{2}/J_{1}$, the degenerate spiral wave vectors cluster around the point~$\mathbf{K}$ ($\mathbf{K}'$), reaching the point~$\mathbf{K}$ ($\mathbf{K}'$) in the limit $J_{2}\gg J_{1}$. Note that in all these phases the spins are in the XY plane. The first correction to the classical limit in the $1/S$ expansion lifts the degeneracy of spiral states and  leaves only spirals on the straight lines $\mathbf{\Gamma}$-$\mathbf{M}$ and $\mathbf{M}$-$\mathbf{K}$, as indicated in Fig.~\ref{fig:1}.

Here, we are rather interested in the ``pure'' spin-1/2 limit of the model~\eqref{eq:XY_model}, where the quantum fluctuations are the strongest and can largely alter the structure of the classical phase diagram. In particular, based on the order-by-disorder analysis, it is known that the quantum regime favors collinear phases with the wave vectors at $\mathbf{\Gamma}$, $\mathbf{M}$, and $\mathbf{K}$ ($\mathbf{K}'$) points. Spiral orders are expected either to completely disappear or to remain only in a small part of the phase diagram. Additionally, completely new phases may arise in the $S=1/2$ regime, with no classical analogs. 

The quantum spin-1/2 model was studied using a plethora of different methods, including variational Monte-Carlo, density matrix renormalization group (DMRG), and coupled cluster methods. At small values of $J_{2}/J_{1}$, the ground state is in the N\'{e}el antiferromagnetic phase. With an increase of $J_{2}/J_{1}$, the model first exhibits a phase transition into either spin liquid or exotic magnetically ordered Ising phase (with magnetization along the $Z$ axis) at around $J_{2}/J_{1} \approx 0.2$. At even higher ratios of $J_{2}/J_{1}$, the model was predicted to host strongly competing valence-bond-solid and collinear magnetic phases. Finally, at $J_{2}/J_{1} \approx 1$ and higher, it was suggested that the model may have a remnant spiral incommensurate magnetic ordering. Note that in this study, we focus on the regime $J_{2}/J_{1} < 1$ and do not discuss the $120$-degree antiferromagnetic (AF) order appearing for $J_{2} \gg J_{1}$.

\section{Method}
We study the phase diagram of the $J_{1}$-$J_{2}$ spin $1/2$ XY model at zero temperature with the variational iPEPS optimization. The general sketch of the algorithm is as follows: we approximate the ground state of the XY model with the iPEPS wave function with a bond dimension $D$, defined on the original honeycomb lattice. The unit cell of the iPEPS wave function consists of two spins on lattice sites A and B, as illustrated in Fig.~\ref{fig:1}. We do not impose additional rotational, reflectional, or continuous U(1) symmetry on the iPEPS wave function. 

To find the best approximation of the ground state with the iPEPS ansatz, we minimize its energy expectation value. 
To optimize the energy, we use a gradient-based approach~\cite{Corboz2016,Vanderstraeten2016}: the energy expectation value is computed approximately with CTMRG~\cite{Nishino1996,Orus2019} on the honeycomb lattice, and its gradients with respect to the iPEPS tensor parameters are obtained via automatic differentiation~\cite{Liao2019}.

In the next subsections, we first introduce our iPEPS ansatz in more detail and then proceed with the discussion of CTMRG generalization, the calculation of observables, and energy optimization. 

\subsection{Spiral iPEPS ansatz}

\begin{figure}
    \includegraphics[width=\linewidth]{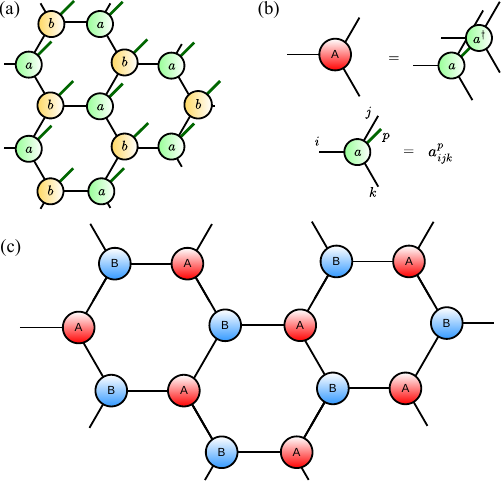}
    \caption{(a) Honeycomb PEPS state with the two-site unit cell consisting of two rank-4 tensors $a$ and $b$.  (b) Definition of two-layer tensor $A$ (the tensor $B$ is defined analogously) necessary to compute the state norm and correlation functions. (c) Infinite tensor network representing the norm of the iPEPS wave function. }
    \label{fig:2}
\end{figure}

As it was previously suggested, both from the classical limit of the model and from the variational Monte-Carlo calculations \cite{DiCiolo2014}, the model can host spiral incommensurate magnetic orders. These types of magnetic ordering cannot be described with a simple iPEPS ansatz, since iPEPS wave functions assume translational invariance with a fixed unit cell. Fortunately, Ref.~\cite{hasik2024incommensurate} proposed a suitable modification of the iPEPS wave functions---dubbed spiral iPEPS---that can be used to describe incommensurate magnetic orders. 

The idea of the modification is to express the wave function as follows: 
\begin{equation}
    |\Psi({\bf q}, a,b)\rangle = U({\bf q})\, |\text{iPEPS}(a,b)\rangle, 
\end{equation}
where $|\text{iPEPS}(a,b)\rangle$ is the usual iPEPS wave function with the translationally invariant unit cell consisting of bulk tensors $a,b$ (as illustrated in Fig.~\ref{fig:2}), and $U({\bf q})$ is the unitary operator, which depends on the incommensurate magnetic order wave vector ${\bf q}$. The unitary transformation factorizes into local operators acting on a single site of the lattice: $U({\bf q}) = \prod_{i} U_{i}({\bf q} \cdot {\bf r}_{i})$, where $U_{i}({\bf q} \cdot {\bf r}_{i})$ acts only on the spin~$i$ and depends on the coordinate ${\bf r}_{i}$ of the unit cell of the site $i$. This operator has the form $U_{i}({\bf q} \cdot {\bf r}_{i}) = \exp{[i ({\bf q} \cdot {\bf r}_{i}) S^{z}_{i}]}$. 

Note that it is possible to define the unitary transformation with the coordinate of the site $i$ inside the unit cell included into its definition, but this part of the unitary transformation, in general, can be absorbed into the iPEPS tensors. The operator rotates the spins in the $XY$ plane by the coordinate-dependent angle, depending on the wave vector ${\bf q}$, which takes values in the Brillouin zone. The tensors $a, b$ in turn have 4 indices: one of dimension $2$ corresponds to the physical on-site spin $1/2$ Hilbert space, while other take values in auxiliary vector spaces of dimension $D$ (also called bond dimension). This bond dimension $D$ is the main parameter controlling the accuracy and expressivity of the iPEPS ansatz. In the numerical calculations, we limit ourselves with the bond dimensions up to $D = 5-6$ in most cases, since the variational optimization of higher-dimensional tensors becomes rather slow. 

The most important property of the unitary transformation $U({\bf q})$ is that for local Hamiltonians $H$, the transformed Hamiltonians $U^{\dagger} H U$ remain local and translationally invariant \cite{hasik2024incommensurate}. Hence, their expectation values with respect to iPEPS wave functions can be calculated.

\subsection{CTMRG}

Exact calculation of observables with the iPEPS wave function is impossible. For this reason, various approximate schemes were developed, such as tensor renormalization group~\cite{Levin2007, Jiang2008}, boundary MPS contraction ~\cite{Jordan2008} or CTMRG \cite{Orus2009, Nishino1996}. In this study, we employ CTMRG on the honeycomb lattice to compute observables and the energy expectation value of the iPEPS wave function. 

The CTMRG on the honeycomb lattice was previously developed in Refs.~\cite{Lukin2023, Nyckees2023}. In this study, we develop a generalization of the honeycomb CTMRG to the more general setting of rotationally non-invariant two-site unit cell honeycomb lattice. Note that another version of anisotropic honeycomb CTMRG was already proposed in Refs.~\cite{Nyckees2023, nyckees2026tensor, ghosh2025simplex, ghosh2025symmetry}. The approach employed here is based on our previous study~\cite{Lukin2024} and is a part of our attempt to generalize CTMRG to different lattices with different unit cells.   In this subsection, we present the steps of this generalized CTMRG. 

\begin{figure}
    \includegraphics[width=\linewidth]{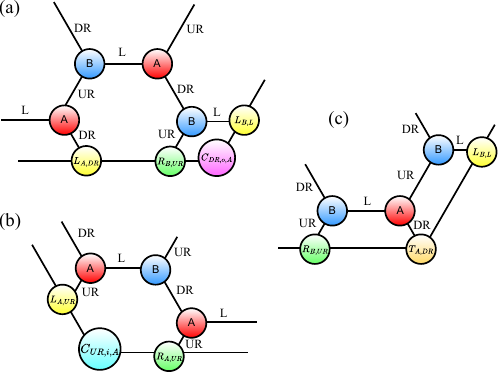} 
    \caption{Definitions of different boundary tensors appearing in the honeycomb CTMRG. The lattice directions are labeled as UR (Upper-Right), DR (Down-Right) and L (Left). All the tensors have also these labels, as well as label $A$ or $B$, which represents the corresponing site tensor. Note that the main bMPS tensors $L$ and $R$ have three types of indices: edge indices connecting them to $A$ or $B$ tensors, indices adjacent to the edge index (with $\pi/3$ angle with edge index direction) and the last non-adjacent index.  (a) $C_{o}$ matrix is placed on the intersection of two bMPS, and its indices are the indices of $L$ and $R$ tensors of these bMPS that are adjacent to the edge leg direction; (b) $C_{i}$ matrices are placed on the intersection of two bMPS, but its indices correspond to the indices of $L$ and $R$ tensors of these bMPS that are not adjacent to the edge leg direction;  (c) $T$ tensors have the same auxiliary indices as $C_{o}$ matrix, but additionally has one edge index. }
    \label{fig:tensors}
\end{figure}

The CTM environment consists of several different types of tensors: bMPS tensors $L, R$, corner tensors $T$ and two different corner matrices $C_{o}$ and $C_{i}$. The tensors $L$ and $R$ (shown in Fig.~\ref{fig:tensors}) have three indices: one edge index of dimension $D^{2}$ and two auxiliary indices of dimension $\chi$. To distinguish these two auxiliary indices we call the index with $\pi/3$ (smaller) angle with respect to the edge index the adjacent index. The corner matrices and corner tensors appear in CTM environments between two bMPS (consisting of $L, R$ tensors), intersecting with $2 \pi /3$ angle, as is illustrated in Fig.~\ref{fig:tensors}. The main difference between $C_{o}$ and $C_{i}$ matrices is that its indices contract with adjacent (non-adjacent) indices of bMPS tensors. Note that all the matrices and tensors have additional labels $A/B$ and direction label $UR/ DR/L$. In case of CTM tensors $L$, $R$ and $T$, the label $A/B$ corresponds to the type of site tensor $A/B$ with which the CTM tensor contracts, while the direction label corresponds to the bond of $A/B$ tensor the CTM tensor contracts with. For the matrices $C_{o}$ and $C_{i}$ the labeling is somewhat more obscure, but generally we label them in the same way as the corresponding tensor $T$ during the CTMRG update step (as discussed below). 

\begin{figure}
    \includegraphics[width=\linewidth]{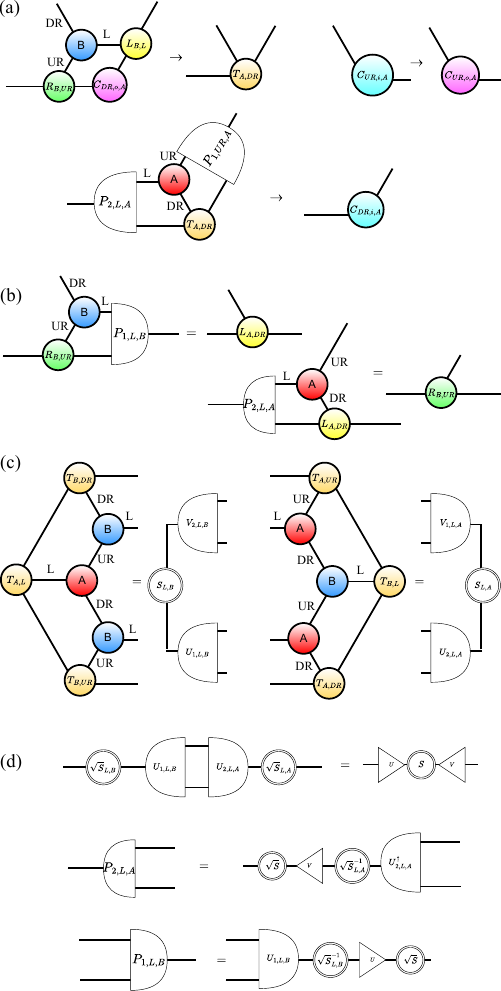} 
    \caption{Illustration of one step of the CTMRG update: (a) new tensors $T$ are obtained from the previous matrices $C_{o}$, new matrix $C_{o}$ is obtained from old $C_{i}$, and new $C_{i}$ is determined by old $T$ (note that $C_{i}$, $C_{o}$, and $T$ are updated simultaneously and not consequently); (b) update of the main bMPS tensors $L$ and $R$ with the introduction of projectors~$P$; (c) to find the projectors $P$, we obtain the half-system environments and then decompose them with SVD decomposition; (d) The SVD tensors from the step (c) are employed to define the new projectors.}
    \label{fig:tensors_update}
\end{figure}

After some initialization (either random of with the site tensors $A, B$) the CTM tensors should be converged to their infinite system values. Convergence is achieved with the repeated application of CTMRG iteration of the update step. These update steps are illustrated in Fig.~\ref{fig:tensors_update}. As shown in Fig.~\ref{fig:tensors_update}(a), the tensors $T$ and corner matrices $C_{o}, C_{i}$ transform into each other. Note that this update of corner matrices and of $T$ tensor is simultaneous in the sense that we use only the $C_{o}, C_{i}, T, L, R$ tensors from the previous step to simultaneously define $C_{o}, C_{i}, T$ tensors of the next step. We also do this update simultaneously for all lattice directions. Fig.~\ref{fig:tensors_update}(b) illustrates the update of bMPS tensors $L$ and $R$ which transform into each other with every iteration. The update of the tensors $L$, $R$, and $T$ employs auxiliary projectors $P$, necessary to truncate the auxiliary bond dimension back to $\chi$. The projectors are found quite similarly to the square lattice CTMRG, as shown in Fig.~\ref{fig:tensors_update}(c) and \ref{fig:tensors_update}(d). 
First, we define the half-lattice environments [shown in Fig.~\ref{fig:tensors_update}(c)] and decompose them with the singular-value decomposition (SVD). Next, the SVD tensors are used to construct projectors. Note that the definition of projectors here is, in fact, equivalent to the 'square-root' variant of CTMRG projectors as defined in the Appendix of Ref. \cite{corboz2014competing}. We used a small regularizing additional constant $\epsilon \approx 10^{-10}$  in the square roots in the projector definitions. 

The CTMRG update steps, as defined above, were used up to CTM convergence, which we measured in terms of convergence of reduced density matrices. During the optimization the CTMRG bond dimensions $\chi$ were chosen slightly higher then $D^{2}$, while in the final observables computation we used $\chi = 2 D^{2}$. 

\subsection{Observables and correlation functions}

To compute observables, it is necessary to find the reduced density matrices of the iPEPS wave function on some neighboring sites. To this end, we  contract the infinite norm tensor network, as shown in Fig.~\ref{fig:2}(c), with physical indices on the chosen neighboring sites remaining uncontracted. For the chosen sites that are geometrically close to each other, the contraction can be performed with CTM environments. 

\begin{figure}
    \includegraphics[width=\linewidth]{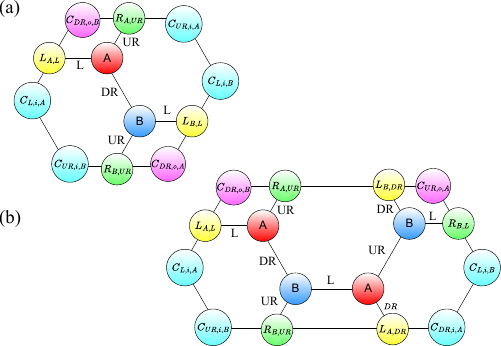}
    \caption{Computation of correlation function with CTM tensors. It can be equivalently used to compute local reduced density matrices, where the physical bra and ket indices of the tensors $A$ and $B$ on the sites relevant for reduced density matrix must be left uncontracted. (a) Computation of the correlation function (or the density matrix) between the two NN sites A and B. (b) Computation of the correlation function between the NN and NNN sites on the honeycomb lattice. }
    \label{fig:correlation}
\end{figure}

In Fig.~\ref{fig:correlation}, we show the contractions of site double layer tensors $A$ and $B$ (with physical indices inside these tensors potentially uncontracted) with CTM tensors necessary to find two-site reduced density matrices between nearest sites, as well as NN and NNN sites. These reduced density matrices can be, in turn, used to compute expectation values of the Hamiltonian of the $J_{1}$-$J_{2}$ model (and potentially of the $J_{1}$-$J_{2}$-$J_{3}$ model as well). 

To find one-site observables, such as magnetization, we  only compute the two-site reduced density matrices and trace over one of the physical indices. 

Note that in contrast to the square lattice CTMRG, for the honeycomb lattice the computation of correlation functions is relatively cheap (with $D^{10}$ scaling in bond dimension, assuming that $\chi \propto D^{2}$). This relatively cheap scaling remains true for even more long-range interactions as well. 

\subsection{Optimization}

To optimize the spiral iPEPS wave function, we implement the CTMRG on the honeycomb lattice with the two-site unit cell and the corresponding correlation functions in the \textsc{PyTorch} automatic differentiation framework \cite{Liao2019, Paszke2019, Hasik2021, pepstorch}. With the loss function defined as the spiral iPEPS energy expectation value, \textsc{PyTorch} allows for the automatic computation of gradients, as the loss computation is defined with only differentiable operations. 

The spiral iPEPS parameters are optimized with the limited-memory Broyden-Fletcher-Goldfarb-Shanno (L-BFGS) algorithm as implemented in \textsc{PyTorch}. To investigate potential competition between different phases, we train the iPEPS parameters starting from different initializations, which are defined either randomly or with some physically inspired initial ansatz (as in the case of dimerized phase wave functions). To increase the variational power of our ansatzes, we also employ wave functions with smaller bond dimensions as the initializations for the wave functions at larger bond dimensions. In this way, the wave functions at larger bond dimensions become close to the energy minima from the start of the optimization.

\section{Results}

There are several open questions that are not completely resolved with the frustrated antiferromagnetic XY~model. The first one is the additional confirmation of the existence of the exotic Ising magnetic phase with the staggered magnetization along the $Z$ direction at intermediate values of $J_{2}/J_{1} \in [0.21, 0.36]$. The second question is the nature of the phase in the interval $J_{2}/J_{1} \in [0.36, 0.5]$, which was previously suggested to be either collinear or dimerized phase. Finally, according to the variational Monte-Carlo calculations and classical model analysis~\cite{DiCiolo2014}, it was suggested that at $J_{2}>J_{1}$ the model may host the spiral magnetic state. The spiral iPEPS ansatz is tailor-made to investigate the potential appearance of this phase. In the next subsections, we discuss our results on these topics separately. The resulting phase diagram is shown in Fig.~\ref{fig:phases}.

\subsection{Transition from N\'{e}el to Ising magnetic order}
\begin{figure}
    \includegraphics[width=\linewidth]{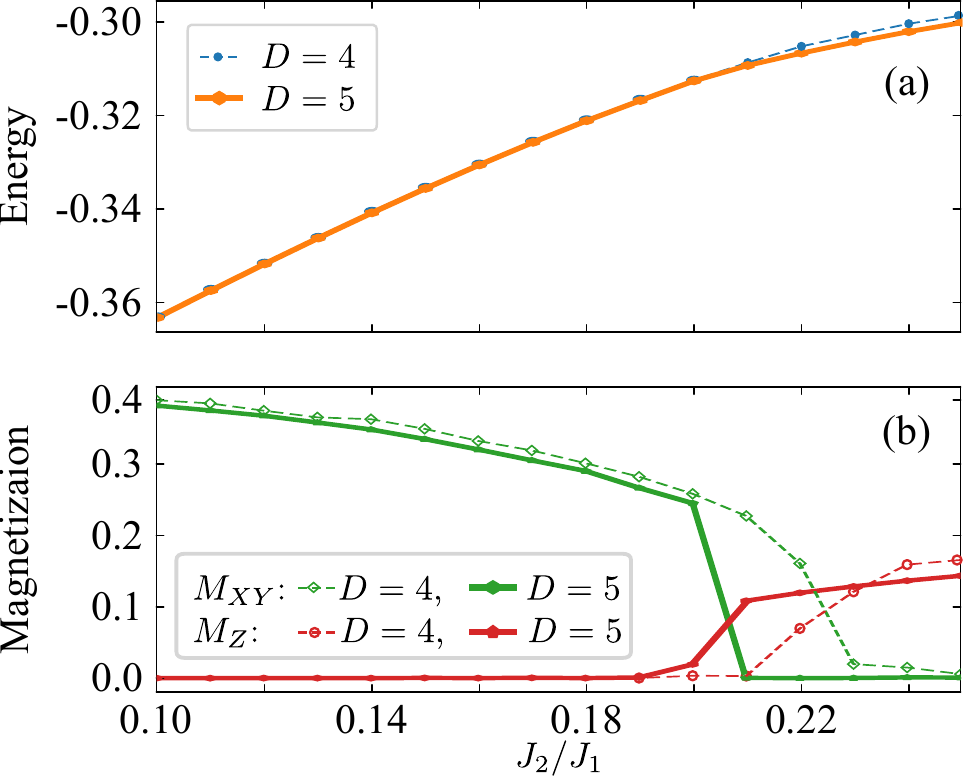}
    \caption{Ground-state energy (a) and {staggered} magnetization in the $XY$-plane $M_{XY}$ and along $Z$ direction $M_{Z}$ (b) as functions of the coupling $J_2$ (in the regime of small to moderate $J_{2}$) indicating phase transition between N\'{e}el and Ising phases. }
    \label{fig:Neel-Ising}
\end{figure}

At $J_2\ll J_1$ the ground state of the model is the N\'{e}el-ordered state with a nonzero staggered magnetization in the $XY$ plane. Our iPEPS analysis confirms this; furthermore, as the ratio $J_2/J_1$ increases, it suggests that the in-plane magnetization continuously decreases to zero while approaching the first critical point $J_2/J_1\approx0.21$. The corresponding behavior of the staggered magnetization and energy is shown in Fig.~\ref{fig:Neel-Ising}.

According to Fig.~\ref{fig:Neel-Ising} we observe that as $J_{2}/J_{1}\to0.21$, the magnetic moments in $XY$-plane are continuously suppressed to zero, similarly to the behavior expected at the second-order phase transition corresponding to the spontaneous breaking of the continuous U(1) spin symmetry. Note that the Ising order at $J_{2}/J_{1}>0.21$ corresponds to the spontaneous breaking of the discrete $Z_2$ symmetry, and it develops only when the magnetic moments in the $XY$ plane  vanish. Note also that the energy does not seem to show any pronounced change in the slope; additionally, the convergence properties of the energy with the bond dimension notably change across the phase transition. This also suggests a second-order phase transition. 


\subsection{Transition from Ising to collinear magnetic order}

At the intermediate values of $J_{2}/J_{1} \in [0.21, 0.36]$, our iPEPS analysis points to the Ising magnetic phase, with spins oriented along the $Z$ axis in antiferromagnetic manner. This phase appearance is highly surprising, since the Hamiltonian does not contain $S^{z} S^{z}$ terms, as first pointed out in Refs.~\cite{zhu2013unexpected, zhu2014quantum} and can be explained by the quantum order-by-disorder arguments~\cite{Oitmaa2014, Jiang2023, chernyshev2025demystifying}.

\begin{figure}
    \includegraphics[width=\linewidth]{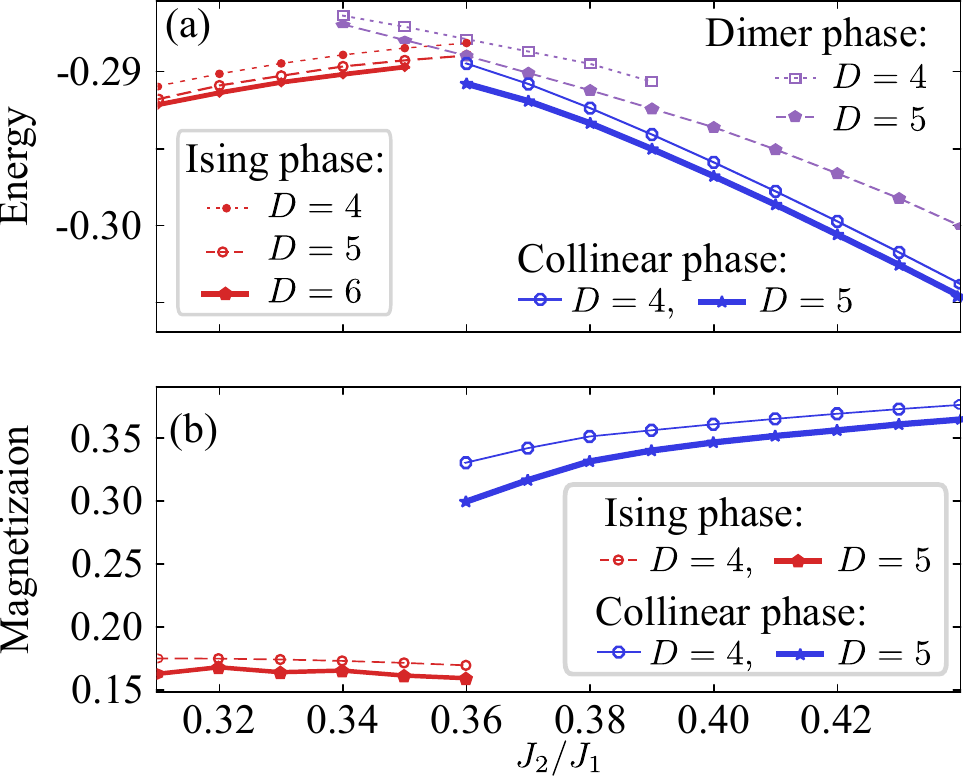}
    \caption{The behavior of energy (a) and average staggered magnetization (b) across the phase transition from the Ising to collinear phase. }
    \label{fig:second_phase_transition}
\end{figure}

In Fig.~\ref{fig:second_phase_transition} we plot the energy and the magnetization in this phase for $J_{2}/J_{1} \in [0.31, 0.36]$. Note that the staggered magnetization is nearly always around $m \approx 0.15$, which is much lower than the mean-field value $m = 0.5$, and is in agreement with the fact that the energy of this state is purely due to quantum fluctuations (as the mean field energy of this phase is zero). The staggered magnetization shows some dependence on the bond dimension~$D$, though this dependence is weak. 

At $J_{2}/J_{1} \approx 0.36$ we observe the phase transition between collinear and Ising phases. The behavior of the energy and magnetization is shown in Fig.~\ref{fig:second_phase_transition}. As the energy clearly changes its slope, we would generally suggest that the transition is of first order. Besides that, it is generally possible to slightly extend PEPS states from the Ising phase to the collinear phase and states from the collinear phase to the Ising phase, in accordance with the expected hysteresis-like behavior, which is another sign of the first-order phase transition.  

\subsection{Competition between dimerized and collinear orders}

In this subsection, our goal is to discuss the two possible phases for $J_{2}/J_{1} \in [0.36, 0.5]$. It was previously observed, both with coupled cluster methods and DMRG, that the dimerized valence bond state and collinear magnetic ordering with the wave vector along in the ${\bf M}$ point of the Brillouin zone strongly compete in this parameter regime \cite{bishop2014frustrated, zhu2014quantum, huang2021quantum}. 

To compare the energies in the two phases, we optimize iPEPS wave functions either from random initial tensors or from dimerized states with Bell-states along some chosen bond of the lattice. With the random initialization, our iPEPS wave functions converged to the collinear phase. Dimerized initial states stayed close to their initial point with some possible very slow drift toward some potential other phase (shown in the form of sometimes slightly nontrivial final wave vector). This suggests that the dimerized state may be at least potentially metastable. Still, in terms of energies the dimerized states were consistently around $1 \%$ higher in energy than the collinear magnetic states. This relation remained valid with a change of state bond dimensions. 

In Fig.~\ref{fig:second_phase_transition} we compare the energies of dimerized and collinear states for different bond dimensions. It can be seen that in all cases the energy of the dimerized state is considerably higher than the energy of collinear state. We should note, that the literature suggested the possibility of mixed collinear and dimerized order. In our calculations we have seen that for $J_{2}/J_{1} \in [0.36, 0.5]$ one bond in the collinear state may have lower energy then the other. We did not see such effect for high $J_{2}/J_{1}$. This energy difference may indicate either some admixture of dimerization to the collinear state or just some artifact of finite bond dimension $D$, as the effect seems to diminish with bond dimension.

\subsection{Spiral incommensurate magnetic order}\label{subsec:spir}

Classically, spirals appear for $J_{2}/J_{1} > 1/6$. Previously, it was reported in Ref.~\cite{DiCiolo2014} that for the quantum $S=1/2$ model, the spirals remain in the parameter region $J_{2}/J_{1} \approx [1, 3.5]$, with the spiral wave vectors located on the straight line between ${\bf M}$ and ${\bf K}$ points in the Brillouin zone (see also Fig.~\ref{fig:1}). Still, it was found that in this region, the spiral states strongly compete with the commensurate orders with ${\bf M}$ or ${\bf K}$ point wavevectors. Note that the results in Ref. \cite{DiCiolo2014} were obtained on the finite clusters, which do not allow dealing directly with arbitrary incommensurate wave vectors and so cannot be totally conclusive. 

The spiral iPEPS ansatz is tailor-made for the study of incommensurate magnetic orders. In particular, as spiral and commensurate magnetic phases at ${\bf M}$ and ${\bf K}$ points are smoothly connected for the spiral iPEPS ansatzes, and as the method works directly in the thermodynamic limit, we can expect that spiral iPEPS will be able to easily discern energy difference between different spirals and commensurate orders. 

According to the performed iPEPS analysis, we observe that the incommensurate magnetic phase stabilizes for $J_{2}/J_{1} > 0.75$ at $D=4$ and for $J_{2}/J_{1} > 0.77$ at $D=5$, which corresponds to much lower values of $J_{2}/J_{1}$ than previously reported. Note that in all cases the variationally optimized wave vector ${\bf q}$ residues precisely on the line between the points ${\bf M}$ and ${\bf K}$ in the Brillouin zone. 

\begin{figure}
    \includegraphics[width=\linewidth]{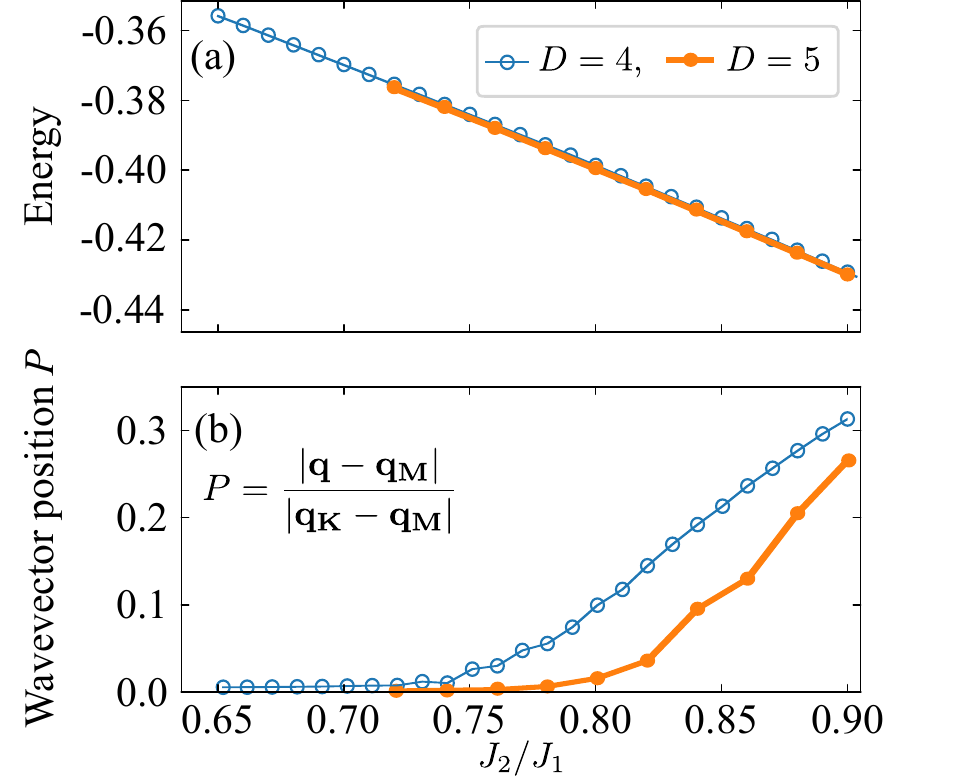}
    \caption{The behavior of the energy (a) and position~$P$ of the spiral wavevector ${\bf q}$ (b) at the transition between collinear and spiral phases. 
    The wavevector ${\bf q}_{\bf M}$ corresponds to the $120$-degree AF phase, which is expected to appear at $J_{2} \gg J_{1}$. }
    \label{fig:spir_phase_transition}
\end{figure}

In Fig.~\ref{fig:spir_phase_transition}(b) we show the dependence of the wave vector~${\bf q}$ on the ratio $J_{2}/J_{1}$. The energy (as shown in Fig.~\ref{fig:spir_phase_transition}(a)) does not show pronounced changes at the transition, which suggests the second-order type. We observe that the spiral phase is generally persistent at higher values of $J_{2}$.

\begin{figure}
    \includegraphics[width=\linewidth]{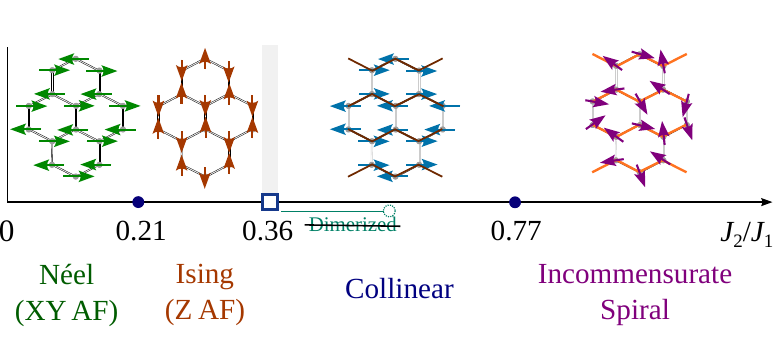}
    \caption{Phase diagram of the $J_1$-$J_2$ antiferromagnetic XY Heisenberg model on the honeycomb lattice according to the iPEPS analysis. The symbols on the horizontal axis indicate the type of the phase transition: the first order (open square) or the second order (filled circles). {Bold colored or thin gray (vertical) lines on the hexagon faces in collinear and spiral phases indicate strong or weak bonds, respectively.}}
    \label{fig:phases}
\end{figure}

In Fig.~\ref{fig:phases}, we also show the general sketch of spin orientations in the spiral phase. We do not include the precise determination of the phase boundary between the spiral phase and the $120$-degree phase in this study, since our calculations for large $J_{2}$ converge to wave vectors close to the ${\bf K}$ point (which corresponds to the $120$-degree order), but they do not reside precisely at it. In this sense, the determination of the precise phase boundary between the two phases is difficult for the given bond dimensions.  

\section{Conclusions and Outlook}

We have developed the CTMRG algorithm on the honeycomb lattice with the two-site unit cell and tested it on the problem of ground state search of the $J_{1}$-$J_{2}$ XY-model on the honeycomb lattice. Since the NNN and next-NNN correlation functions can be computed with the same computational cost in the framework of the honeycomb CTMRG, as the NN correlators, the models with long-range interactions are not more complicated for the detailed study, than the models with only nearest-neighbor interactions. For the ground-state search, we developed and employed the spiral iPEPS tensor network ansatz, which allows for the treatment of spiral phases, previously predicted for the model with the variational Monte Carlo approach. 

The obtained phase diagram, as shown in Fig.~\ref{fig:phases}, is generally in agreement with the previous studies on the subject: we find the N\'{e}el phase at low $J_{2}/J_{1}$, with a transition to exotic Ising order at $J_{2}/J_{1} \approx 0.21$. At yet higher $J_{2}/J_{1} \approx 0.36$, we observe the first-order phase transition to the collinear phase. Note that the valence-bond-solid phase (predicted in some previous studies) can be obtained with iPEPS optimization, but in all the cases its energy appears higher than the energy of the collinear phase. At higher $J_{2}/J_{1} \approx 0.8$, we find the transition to the spiral magnetic phase with the wave vector ${\bf q}$ on the {\bf M}-{\bf K} line in the Brillouin zone. 

Further research in this direction can be devoted to the development of CTMRG algorithms with nontrivial unit cells on different lattices and their further application to iPEPS optimization problems. 
As for the honeycomb lattice, the next research direction could be either the XY model with a third-neighbor coupling ($J_{1}$-$J_{2}$-$J_{3}$ XY model)~\cite{zhu2014quantum} or the Heisenberg $J_{1}$-$J_{2}$-$J_{3}$ counterpart~\cite{Fouet2001, Oitmaa2011}.

\acknowledgements
The authors acknowledge support by the National Research Foundation of Ukraine, project No.~2023.03/0073.

\bibliography{references}
\end{document}